\documentclass[aps,12pt,a4paper,showpacs,byrevtex]{revtex4}
\usepackage{amsmath}
\usepackage{bm}
\usepackage{amsfonts}
\begin{document}
\baselineskip 15pt

\title{Geometric phases for wave packets in a uniform magnetic
field}\thanks{published in Phys. Lett. A {\bf 298} (2002) 67-72.}



\author{Qiong-Gui Lin}
\email[]{qg_lin@163.net}


\affiliation{China Center of Advanced Science and Technology (World
Laboratory), P. O. Box 8730, Beijing 100080, People's Republic of
China}
\thanks{not for correspondence}
\affiliation{Department of Physics, Sun Yat-Sen University, Guangzhou
510275, People's  Republic of China}


\begin{abstract}
\baselineskip 15pt {\normalsize A wave packet of a charged particle
always make cyclic circular motion in a uniform magnetic field, just
like a classical particle. The nonadiabatic geometric phase for an
arbitrary wave packet can be expressed in terms of the mean value of
a number operator. For a large class of wave packets, the geometric
phase is proportional to the magnetic flux encircled by the orbit of
the wave packet. For more general wave packets, however, the
geometric phase contains an extra term.}
\end{abstract}

\maketitle



\section{\label{s1} Introduction}

Since the discovery of the geometric phase \cite{berry,aha,sam}, the
subject has been studied for various systems
\cite{li-book,wu-li,jordan,chen-ni,berry-jmo}, especially particles
with spin in time-dependent magnetic fields
\cite{wang,wagh1,wagh2,fernandez,layton,gao,ni,zhu,pra01,jpa01,jpa02}.
To our knowledge the geometric phases for wave packets that undergo
cyclic motion in the configuration space were not considered in the
literature. Of special interest is the wave packets of a charged
particle moving in a uniform magnetic field, because this represents
a rather simple and realizable physical situation. It should be noted
that a geometric phase can occur even though the system is time
independent, just like spin in a uniform magnetic field \cite{aha}.

In this paper we consider a charged particle moving in a uniform
magnetic field. In Sec. \ref{s2} we briefly review the energy levels
and the eigenstates of the system. Though the solutions to the system
is well known in the textbooks \cite{landau,ni-book}, we present the
results somewhat differently. In Sec. \ref{s3} we consider the time
evolution of an arbitrary wave packet. The time evolution is always
cyclic and the center of the wave packet makes circular motion like a
classical particle regardless of the initial condition. The
nonadiabatic geometric phase in a cycle can be expressed in terms of
the mean value of a number operator. This is similar to the case for
a particle with spin in a rotating magnetic field, where the
nonadiabatic geometric phase can be expressed in terms of the mean
value of the component of the angular momentum along the rotating
axis \cite{wang,pra01,jpa01}. We also establish a linear relation
between the nonadiabatic geometric phase and the magnetic flux
encircled by the circular orbit of the wave packet. There is an
inhomogeneous term in this relation, which we call an extra term. A
large class of wave packet is presented in Sec. \ref{s4} for which
the extra term vanishes and thus the nonadiabatic geometric phase is
proportional to the magnetic flux. For wave packets that do not
belong to this class, the extra term depends on the initial
condition. Again this is similar to the case for particles with spin
in a rotating magnetic field \cite{jpa02}. A brief summary is given
in Sec. \ref{s5}.

\section{\label{s2} Review of the model}

Consider a charged particle with electric charge $q$ and mass $M$,
moving on the $xy$ plane under the influence of a uniform magnetic
field $\mathbf B=B\mathbf e_z$ where $\mathbf e_z$ is the unit vector
in the $z$ direction and $B$ is a constant which can be taken as
positive without loss of generality. We take the symmetric gauge
$A_x=-By/2$, $A_y=Bx/2$, then the Hamiltonian reads
\begin{equation}\label{1}
H={1\over 2M}(p_x^2+p_y^2)+{1\over 8}M\omega_B^2(x^2+y^2)-\frac 12
\epsilon\omega_B L_z,
\end{equation}
where $\omega_B=|q|B/Mc>0$, $L_z=xp_y-yp_x$, and $\epsilon=1$ ($-1$)
if $q$ is positive (negative). We define
\begin{equation}\label{2}
a_1={M\omega_B x+\mathrm i 2p_x\over2\sqrt{M\hbar\omega_B}},\quad
a_2={M\omega_B y+\mathrm i 2p_y\over2\sqrt{M\hbar\omega_B}}.
\end{equation}
The nonvanishing commutators among these operators and their
Hermitian conjugates are $[a_i, a_j^\dag]=\delta_{ij}$. The
Hamiltonian can be recast as
\begin{equation}\label{3}
H=\frac12 \hbar\omega_B[(a_1^\dag a_1+a_2^\dag a_2+1)+\mathrm
i\epsilon (a_1^\dag a_2-a_1 a_2^\dag)].
\end{equation}
Next we define
\begin{equation}\label{4}
a={1\over\sqrt 2}(a_1+\mathrm i\epsilon a_2),\quad b={1\over\sqrt 2}
(a_1-\mathrm i\epsilon a_2),
\end{equation}
and their Hermitian conjugates. They satisfy
\begin{equation}\label{5}
[a,a^\dag]=[b,b^\dag]=1,
\end{equation}
and all other commutators vanish. In terms of these operators, we
have
\begin{equation}\label{6}
H=\hbar\omega_B(a^\dag a+\textstyle\frac12),
\end{equation}
and
\begin{equation}\label{7}
L_z=\epsilon\hbar(b^\dag b-a^\dag a).
\end{equation}

We see that the Hamiltonian becomes that for a simple harmonic
oscillator. Thus the energy levels are
\begin{equation}\label{8}
E_n=\hbar\omega_B(n+\textstyle\frac12),\quad n=0,1,2,\ldots.
\end{equation}
One can take the eigenstates to be the common ones of $N=a^\dag a$
and $N'=b^\dag b$, denoted by $|nn'\rangle$, satisfying
\begin{equation}\label{9}
N|nn'\rangle=n|nn'\rangle,\quad N'|nn'\rangle=n'|nn'\rangle,\quad
n,n'=0,1,2,\ldots.
\end{equation}
They are also common eigenstates of $H$ and $L_z$, with eigenvalues
$E_n$ and $\epsilon(n'-n)\hbar$, respectively. These states are given
by
\begin{equation}\label{10}
|nn'\rangle={1\over\sqrt{n!n'!}}(a^\dag)^n (b^\dag)^{n'}|00\rangle,
\end{equation}
where the ground state $|00\rangle$ satisfies
\begin{equation}\label{11}
a|00\rangle=b|00\rangle=0.
\end{equation}
One can work out the wave functions in the configuration space for
these eigenstates and show that they are essentially the same as
those obtained by solving the Schr\"odinger equation in the
cylindrical coordinates. However, we are not interested in the
quantum number $n'$ in this paper, so we will only deal with the
quantum number $n$. We consider the eigenstates $|n\rangle$ of $N$ or
$H$, satisfying
\begin{equation}\label{12}
N|n\rangle=n|n\rangle, \quad H|n\rangle=E_n|n\rangle, \quad
n=0,1,2,\ldots.
\end{equation}
These are also called number states in the following. They are given
by
\begin{equation}\label{13}
|n\rangle={1\over\sqrt{n!}}(a^\dag)^n |0\rangle,
\end{equation}
where the ground state $|0\rangle$ satisfies
\begin{equation}\label{14}
a|0\rangle=0.
\end{equation}
Obviously, the state $|n\rangle$ is a linear combination of
$|nn'\rangle$, that is
\begin{equation}\label{15}
|n\rangle=\sum_{n'=0}^\infty C_{n'}|nn'\rangle,
\end{equation}
where the coefficients $C_{n'}$ are arbitrary except satisfying
$\sum_{n'=0}^\infty |C_{n'}|^2=1$ such that $|n\rangle$ is
normalized. Therefore there must exist some big freedom in the wave
function for the state $|n\rangle$.

We define a complex number $z$ and its complex conjugate $z^*$ as
\begin{equation}\label{16}
z=\sqrt{M\omega_B\over 4\hbar}(x+\mathrm i\epsilon y),\quad
z^*=\sqrt{M\omega_B\over 4\hbar}(x-\mathrm i\epsilon y),
\end{equation}
then in the configuration space we have
\begin{equation}\label{17}
a={1\over\sqrt2}(z+\partial_{z^*}),\quad a^\dag={1\over\sqrt2}(
z^*-\partial_{z}).
\end{equation}
The wave function for the ground state $|0\rangle$ is obviously
\begin{equation}\label{18}
\psi_0(z,z^*)=\exp(-z^*z)f(z),
\end{equation}
where the function $f(z)$ is such that $\psi_0(z,z^*)$ is well
behaved everywhere and is normalizable. Thus $f(z)$ is rather
arbitrary. In particular, any polynomial satisfies the requirement.
We assume that $\psi_0(z,z^*)$ has been normalized, then the
normalized wave function for the higher excited state $|n\rangle$ is
\begin{equation}\label{19}
\psi_n(z,z^*)={(-)^n\over\sqrt{2^n n!}} \exp(z^*z) \partial_z^n
[\exp(-2z^*z)f(z)].
\end{equation}
Therefore the freedom in these eigenstates lies in the arbitrariness
of $f(z)$. Of course this freedom corresponds to the degeneracy of
the Landau energy levels (\ref{8}). An arbitrary state of the system
can be expressed as a linear combination of the above number states.

A class of states (or wave packets in the configuration space) that
are of special interest in the following are the so called displaced
number states. Similar to those for a simple harmonic oscillator
\cite{ni-book,ka,nieto}, they are defined as
\begin{equation}\label{20}
|n,\alpha\rangle=D(\alpha)|n\rangle,
\end{equation}
where $D(\alpha)$ is the unitary displacement operator
\begin{equation}\label{21}
D(\alpha)=\exp(\alpha a^\dag-\alpha^* a),
\end{equation}
where $\alpha$ is a complex number. One can define a more general
displaced state $|\varphi,\alpha \rangle$ by acting $D(\alpha)$ on an
arbitrary state $|\varphi\rangle$. An important property is that if
the wave function for $|\varphi\rangle$ is $\varphi(z,z^*)$ or
$\varphi(x,y)$, then that for $|\varphi,\alpha \rangle$ is
\begin{subequations}\label{22}
\begin{equation}\label{22a}
\varphi_\alpha(z,z^*)=\exp\left({\alpha z^*-\alpha^*
z\over\sqrt2}\right)\varphi\left(z-{\alpha\over\sqrt2}, z^*-
{\alpha^*\over\sqrt2}\right),
\end{equation}
or
\begin{equation}\label{22b}
\varphi_\alpha(x,y)=\exp\left[\mathrm i\sqrt{M\omega_B\over 2\hbar}
(\alpha_y x-\epsilon\alpha_x y) \right]
\varphi\left(x-\sqrt{2\hbar\over M\omega_B}\alpha_x,
y-\epsilon\sqrt{2\hbar\over M\omega_B} \alpha_y \right),
\end{equation}
\end{subequations}
where $\alpha_x=\mathrm{Re}\,\alpha$, $\alpha_y=\mathrm{Im}\,\alpha$.
Thus the wave packet of the displaced state is essentially an entire
displacement of the original one, and the above definition for the
displacement operator $D(\alpha)$ seems to be sound.

\section{\label{s3} The geometric phase for an arbitrary wave packet}

Consider the time evolution of an arbitrary wave packet. The initial
state at $t=0$ may be expressed as
\begin{equation}\label{23}
|\psi(0)\rangle=\sum_{n=0}^\infty c_n|n\rangle.
\end{equation}
where the coefficients $c_{n}$ are arbitrary except satisfying
$\sum_{n=0}^\infty |c_{n}|^2=1$ such that $|\psi(0)\rangle$ is
normalized. Since the Hamiltonian is given by Eq. (\ref{6}), the
state at a latter time $t$ is
\begin{equation}\label{24}
|\psi(t)\rangle=\mathrm e^{-\mathrm i Ht/\hbar}|\psi(0)\rangle
=\sum_{n=0}^\infty c_n\exp[-\mathrm i\omega_B t(n+\textstyle\frac12)]
|n\rangle.
\end{equation}
At the time $T=2\pi/\omega_B$, we have
\begin{equation}\label{25}
|\psi(T)\rangle=\mathrm e^{-\mathrm i \pi}|\psi(0)\rangle.
\end{equation}
Therefore any state is cyclic, and the total phase change in a cycle
is
\begin{equation}\label{26}
\delta=-\pi, \quad \mathrm{mod}~2\pi,
\end{equation}
which is independent of the initial condition. The expectation value
of $H$ in the state $|\psi(t)\rangle$ is
\begin{equation}\label{27}
\langle H\rangle_t\equiv \langle \psi(t)|H|\psi(t)\rangle =\langle
\psi(0)|H|\psi(0)\rangle=\hbar\omega_B(\langle N\rangle+\textstyle
\frac12),
\end{equation}
where the mean value $\langle N\rangle$ is evaluated in the initial
state. The dynamic phase is
\begin{equation}\label{28}
\beta=-\hbar^{-1}\int_0^T \langle H\rangle_t\, \mathrm dt =-\pi
-2\pi\langle N\rangle.
\end{equation}
This depends on the initial state. The nonadiabatic geometric phase
is
\begin{equation}\label{29}
\gamma=\delta-\beta=2\pi\langle N\rangle, \quad \mathrm{mod}~2\pi.
\end{equation}
It is proportional to the mean value of the number operator $N$. This
is similar to the case for a particle with spin in a rotating
magnetic field, where the nonadiabatic geometric phase can be
expressed in terms of the mean value of the component of the angular
momentum along the rotating axis \cite{wang,pra01,jpa01}. A number
state has obviously a vanishing geometric phase (modulo $2\pi$) since
its time evolution is trivial.

On the other hand, we consider the motion of the center of the wave
packet. The position of it is characterized by the mean value of the
coordinate variables $x$ and $y$. We denote
\begin{equation}\label{30}
\beta_a=\langle\psi(0)|a|\psi(0)\rangle, \quad
\beta_b=\langle\psi(0)|b|\psi(0)\rangle.
\end{equation}
Using the relations
\begin{equation}\label{31}
\mathrm e^{\mathrm iHt/\hbar}a\mathrm e^{-\mathrm iHt/\hbar}
=a\exp(-\mathrm i\omega_B t),\quad \mathrm e^{\mathrm iHt/\hbar}
b\mathrm e^{-\mathrm iHt/\hbar} =b,
\end{equation}
which can be easily verified, we have
\begin{equation}\label{32}
\langle\psi(t)|a|\psi(t)\rangle=\beta_a\exp(-\mathrm i\omega_B t),
\quad \langle\psi(t)|b|\psi(t)\rangle=\beta_b.
\end{equation}
This leads to
\begin{subequations}\label{33}
\begin{equation}\label{33a}
\bar x_t-\sqrt{2\hbar\over M\omega_B}\beta_{bx}=\sqrt{2\hbar\over
M\omega_B}|\beta_a|\cos(\omega_B t-\arg\beta_a),
\end{equation}
\begin{equation}\label{33b}
\bar y_t+\epsilon\sqrt{2\hbar\over M\omega_B}\beta_{by}
=-\epsilon\sqrt{2\hbar\over M\omega_B}|\beta_a|\sin(\omega_B
t-\arg\beta_a),
\end{equation}
\end{subequations}
where $\bar x_t=\langle\psi(t)|x|\psi(t)\rangle$, $\bar y_t =\langle
\psi(t) |y|\psi(t)\rangle$, and $\beta_{bx}=\mathrm{Re}\,\beta_b$,
$\beta_{by}=\mathrm{Im}\,\beta_b$. Therefore the center of the wave
packet always makes a circular motion. The angular frequency is
$\omega_B$. The motion is clockwise when $q>0$ and anticlockwise when
$q<0$. These are all the same as for a classical particle. The radius
of the circle depends only on $\beta_a$, while $\beta_b$ only
determines the position of the center of the circle. The magnetic
flux go through the area encircled by the circular orbit is defined
as positive (negative) if the motion is anticlockwise (clockwise).
This turns out to be
\begin{equation}\label{34}
\Phi=-{\hbar c\over q}2\pi|\beta_a|^2 =-{\hbar c\over q}2\pi\langle
a^\dag\rangle\langle a\rangle=-{\hbar c\over q}2\pi|\langle
a\rangle|^2,
\end{equation}
where the expectation values are evaluated in the initial state.
Compared with Eq. (\ref{29}) we obtain
\begin{equation}\label{35}
\gamma=-{q\Phi\over \hbar c}+2\pi(\Delta a)^2, \quad
\mathrm{mod}~2\pi.
\end{equation}
where
\begin{equation}\label{36}
(\Delta a)^2\equiv \langle a^\dag a\rangle- \langle a^\dag\rangle
\langle a\rangle =\langle a^\dag a\rangle- |\langle a\rangle|^2
=\langle (a-\langle a\rangle)^\dag(a-\langle a\rangle)\rangle.
\end{equation}
Thus the nonadiabatic geometric phase contains two terms, the first
is proportional to the magnetic flux encircled by the orbit of the
center of the wave packet, the second is an ``extra term''. In Sec.
\ref{s4} we will show that the extra term vanishes (modulo $2\pi$) if
the wave packet is initially a displaced number state. For a more
general wave packet, however, the extra term depends on the initial
condition. Again this is similar to the case for particles with spin
in a rotating magnetic field \cite{jpa02}.

\section{\label{s4} The geometric phase for a displaced number state}

Consider the special case where the initial state is a displaced
number state
\begin{equation}\label{37}
|\psi(0)\rangle=|n,\alpha\rangle=D(\alpha)|n\rangle.
\end{equation}
Using the relations
\begin{equation}\label{38}
D^\dag(\alpha)aD(\alpha)=a+\alpha, \quad D^\dag(\alpha)a^\dag
D(\alpha)=a^\dag+\alpha^*,
\end{equation}
which can be easily verified, we have
\begin{equation}\label{39}
\langle a\rangle=\langle \psi(0)|a|\psi(0)\rangle=\alpha, \quad
\langle a^\dag\rangle=\langle\psi(0) |a^\dag| \psi(0)\rangle
=\alpha^*,
\end{equation}
and
\begin{equation}\label{40}
\langle a^\dag a\rangle=\langle \psi(0)|a^\dag
a|\psi(0)\rangle=n+|\alpha|^2.
\end{equation}
Therefore in these states
\begin{equation}\label{41}
(\Delta a)^2=n,
\end{equation}
and
\begin{equation}\label{42}
\gamma=-{q\Phi\over \hbar c}, \quad \mathrm{mod}~2\pi.
\end{equation}
This holds for all numbers $n=0,1,2,\ldots$, and for any complex
number $\alpha$. Once again this result is similar to the case for
spin in a time-dependent magnetic field where for special initial
conditions the nonadiabatic geometric phase is always proportional to
the solid angle subtended by the trace of the spin vector. (For a
rotating magnetic field this is well known. For an arbitrarily
varying magnetic field, see \cite{layton,gao,jpa02}.)

It should be noticed that the time evolution of a displaced number
state is rather simple. Using relations similar to Eq. (\ref{31}), it
is easy to show that
\begin{equation}\label{43}
\mathrm e^{-\mathrm iHt/\hbar}D(\alpha)\mathrm e^{\mathrm iHt/\hbar}
=D(\alpha_t),
\end{equation}
where
\begin{equation}\label{44}
\alpha_t=\alpha\exp(-\mathrm i\omega_B t).
\end{equation}
Thus for the initial condition (\ref{37}), the time evolution is
given by
\begin{equation}\label{45}
|\psi(t)\rangle=\exp[-\mathrm i\omega_B
t(n+\textstyle\frac12)]D(\alpha_t) |n\rangle =\exp[-\mathrm i\omega_B
t(n+\textstyle\frac12)] |n,\alpha_t\rangle.
\end{equation}
We see that it remains to be a displaced number state at all latter
times, except that the displacement parameter varies with time, and
an overall phase factor is gained. Since the displacement only
changes the position of the wave packet but not the shape of it, we
conclude that a displaced number state keeps its shape unchanged
while it is making circular motion.

\section{\label{s5} Summary}

In this paper we have studied the time evolution and the associated
nonadiabatic geometric phase for a wave packet of a charged particle
moving in a uniform magnetic field. Any state of this system is
cyclic, and the center of it always makes circular motion like a
classical particle. The nonadiabatic geometric phase in a cycle can
be expressed in terms of the mean value of a number operator. A
linear relation between the nonadiabatic geometric phase and the
magnetic flux encircled by the circular orbit of the wave packet is
established. For wave packets that are initially displaced number
states the nonadiabatic geometric phase is proportional to the
magnetic flux. For more general wave packets it contains an extra
term which depends on the initial condition. This shows that the
nonadiabatic geometric phase is not always proportional to the
geometric object (the magnetic flux in the present problem is a
geometric object since it is proportional to the area through with it
penetrates) involved in the cyclic motion, but may be a more general
function of the latter. Similar results were encountered in the
problem of particles with spin moving in a rotating magnetic field
\cite{jpa02}. We also showed that a displaced number state keeps its
shape unchanged while it is making circular motion.

\begin{acknowledgments}
This work was supported by the National Natural Science Foundation of
the People's Republic of China, and by the Foundation of the Advanced
Research Center of Sun Yat-Sen University.
\end{acknowledgments}


\end{document}